\documentclass{jps-cp}
\usepackage{txfonts}
\def\eff{{\rm eff}}
\def\AF{{\rm AF}}
\def\F{{\rm F}}
\def\P{{\rm P}}
\title{Nematic Tomonaga-Luttinger Liquid Phase in an
\mbox{\boldmath $S\!=\!1/2$} Ferromagnetic-Antiferromagnetic Bond-Alternating
Chain}

\author{Takashi \textsc{Tonegawa}$^{1,2,3}$,
        Kiyomi \textsc{Okamoto}$^{2}$,
        Kiyohide \textsc{Nomura}$^{4}$
        and T\^oru \textsc{Sakai}$^{2,5}$}

\inst{$^{1}$ Professor Emeritus, Kobe University, Kobe 657-8501, Japan\\
$^{2}$ Graduate School of Science, University of Hyogo, Hyogo
678-1297, Japan\\
$^{3}$ Department of Physics, Osaka Metropolitan University, Sakai 599-8531,
Japan\\
$^{4}$ Department of Physics, Kyushu University, Fukuoka 812-8581, Japan\\
$^{5}$ National Institute for Quantum Science and Technology (QST), SPring-8,
Hyogo 679-5148, Japan}

\email{tone0115@vivid.ocn.ne.jp}

\recdate{June 29, 2022}

\abst{We numerically investigate the ground-state phase diagram of the
\hbox{$S\!=\!1/2$} ferromagnetic-antiferromagnetic bond-alternating
chain, in which the ferromagnetic interactions are stronger than the
antiferromagnetic ones, and the anisotropies of the former and latter
interactions are of the Ising-type and the $XY$-type, respectively.  We use
various numerical methods, such as the level spectroscopy and phenomenological
renormalization-group analyses of the numerical data obtained by the exact
diagonalization method, and so on.  The resultant phase diagrams contain the
ferromagnetic, $XY$1, singlet-dimer, and up-up-down-down phases as well as the
nematic Tomonaga-Luttinger liquid (nTLL) phase which appears in a wide region
of the interaction parameters.  Perturbation calculations from the strong limit
of the ferromagnetic interactions reproduce fairly well the numerical results
of the phase boundary lines associated with the nTLL phase in the phase
diagrams.}

\kword{\hbox{$S\!=\!1/2$} ferromagnetic-antiferromagnetic bond-alternating
       chain, ground-state phase diagram, nematic Tomonaga-Luttinger liquid
       phase, numerical calculation, perturbation theory}

\begin{document}
\maketitle

\section{Introduction}

Over the past several decades, a great deal of numerical, theoretical, and
experimental work has been devoted to clarifying the quantum phase transition
in one-dimensional \hbox{$S\!=\!1/2$} systems.  As is well known, due to the
strong quantum fluctuation, a variety of exotic quantum phases appear in the
ground-state phase diagrams of these systems.  A typical example attracted
recently much attention is the
nematic Tomonaga-Luttinger liquid (nTLL) phase~\cite{chubukov,vekua,hikihara-1,
sudan,sakai-1,sato,starykh,buttgen,orlova,parvej,mendels,tonegawa-1}, which is
characterized not only by the formation of two-magnon bound pairs but also by
the dominant nematic four-spin correlation function.

Recently, we\cite{tonegawa-1} have  numerically explored the ground-state phase
diagram of an \hbox{$S\!=\!1/2$} anisotropic two-leg ladder with different
leg interactions in the absence of external magnetic field.  This system has
frustration when the signs of the two kinds of leg interactions
are different from each other.  Then, we have found that, when the
ferromagnetic rung interactions with the Ising-type anisotropy is much stronger
than the antiferromagnetic leg interactions with the $XY$-type anisotropy, the
nTLL phase appears in the unfrustrated region as well as in the frustrated
region.  For this phase in the former region, the asymptotic form of the
nematic correlation function shows the power-law decay with the uniform
character; on the other hand, for this phase in the latter region, that
shows the power-law decay with the staggered character.  Thus, both nTLL
phases are different phases.  As far as we know, this is the first report of
the realization of the nTLL phase in an \hbox{$S\!=\!1/2$} unfrustrated
one-dimensional spin system under no external magnetic field.

According to this result, it is reasonably expected that the nTLL state appears
as the zero-field ground state in general \hbox{$S\!=\!1/2$} unfrustrated
one-dimensional systems in which pairs of \hbox{$S\!=\!1/2$} spins coupled
strongly with the Ising-type ferromagnetic interaction are connected by the
weak antiferromagnetic interactions.  As an example of such systems, we
investigate in this paper an \hbox{$S\!=\!1/2$} ferromagnetic-antiferromagnetic
bond-alternating chain.  We express the Hamiltonian describing this system as
\begin{equation}
    {\cal H}
       = - J_{{\rm F}} \sum_{j=1}^{N/2}
              \bigl\{\Gamma_{{\rm F}} 
              \bigl(S_{2j-1}^x S_{2j}^x + S_{2j-1}^y S_{2j}^y\bigr)
                     + S_{2j-1}^z S_{2j}^z \bigr\}
         + J_{\rm AF} \sum_{j=1}^{N/2}
              \bigl\{S_{2j}^x S_{2j+1}^x + S_{2j}^y S_{2j+1}^y
                           + \Delta_{{\rm AF}}\,S_{2j}^z S_{2j+1}^z \bigr\}\,,
\label{eq:hamiltonian}
\end{equation}
where  $S_j^\mu$ (\hbox{$\mu\!=\!x,y,z$}) is the $\mu$-component of the
\hbox{$S\!=\!1/2$} operator \mbox{\boldmath $S$}$_j$ at the $j$th site;
$J_{{\rm F}}$ and $J_{{\rm AF}}$ denote, respectively, the magnitudes of the
ferromagnetic and antiferromagnetic interactions; $\Gamma_{{\rm F}}$ and
$\Delta_{{\rm AF}}$ are, respectively, the parameters representing the
$XXZ$-type anisotropies of the former and latter interactions; $N$ is the
total number of spins in the system, which is assumed to be a multiple of
four.  We assume that \hbox{$J_{{\rm F}}\!>\!J_{{\rm AF}}\!>\!0.0$},
$1.0 \geq \Gamma_{{\rm F}} \ge 0.0$, and $1.0 \ge |\Delta_{{\rm AF}}|$,
that is, the ferromagnetic interactions are stronger than the 
antiferromagnetic ones, and the anisotropies of the former and latter
interactions are of the Ising-type and the $XY$-type, respectively.

\section{Ground-State Phase Diagram}

\begin{figure}[t]
   \begin{center}
      \includegraphics[scale=0.29]{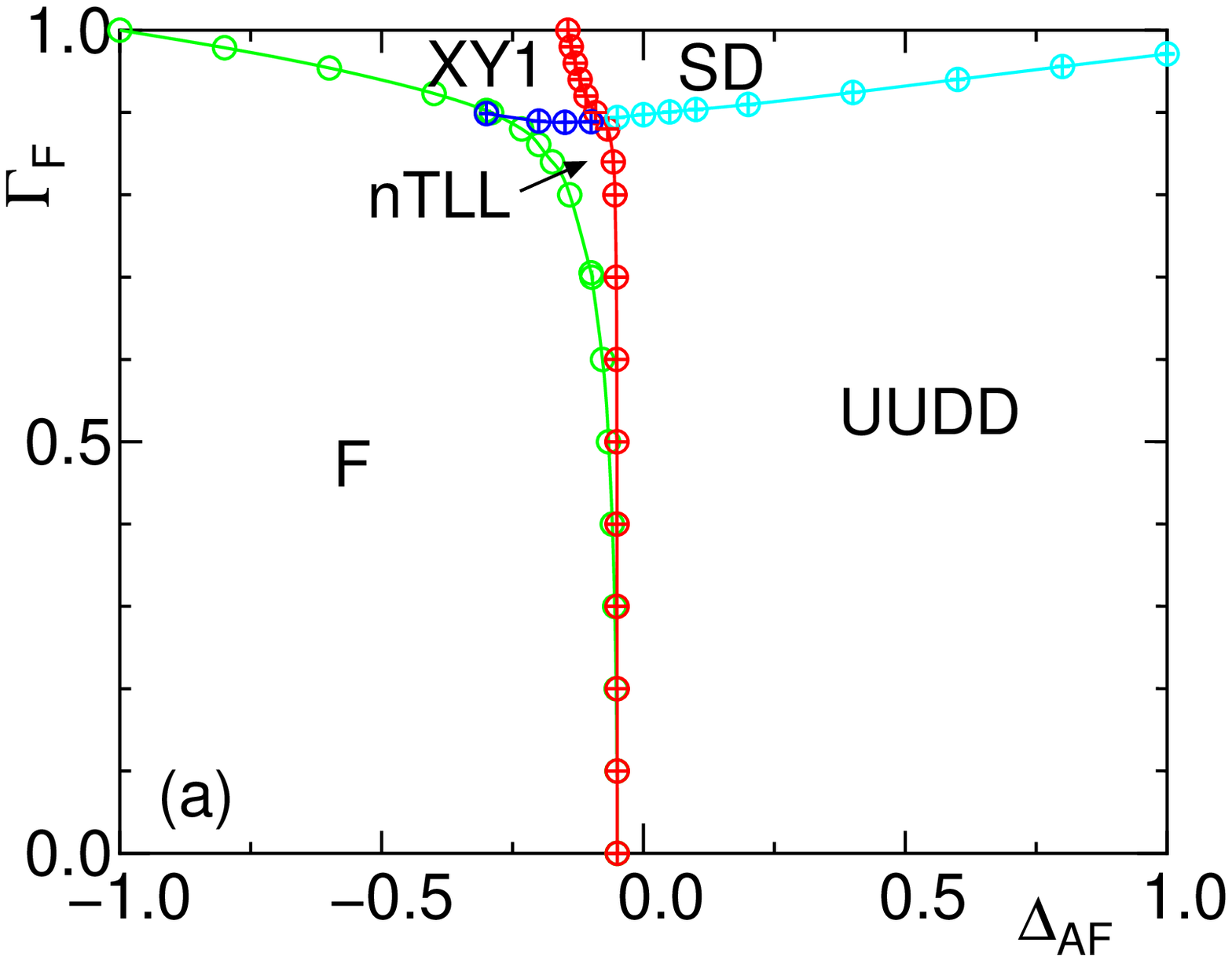}~~
      \includegraphics[scale=0.29]{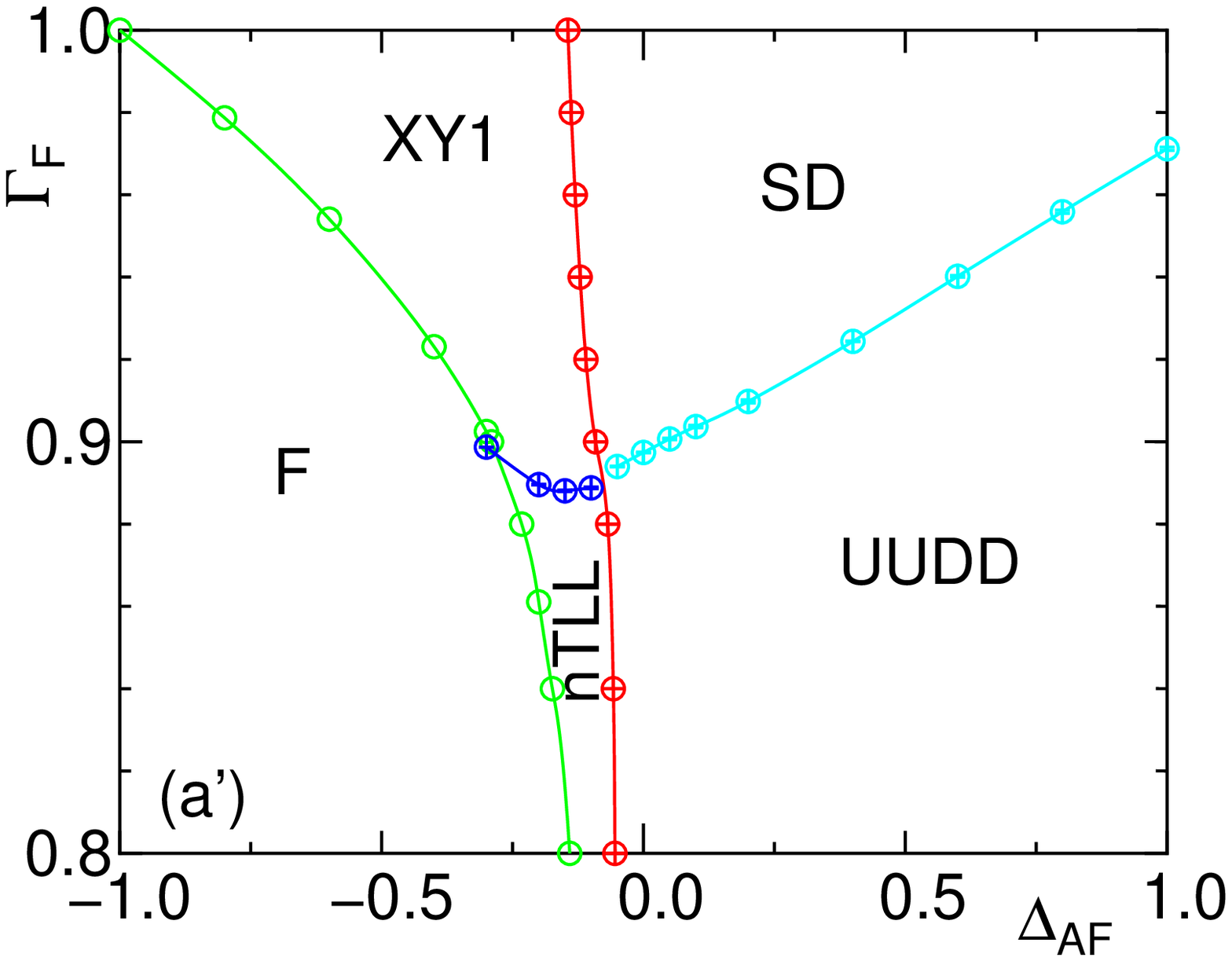}~~
      \includegraphics[scale=0.29]{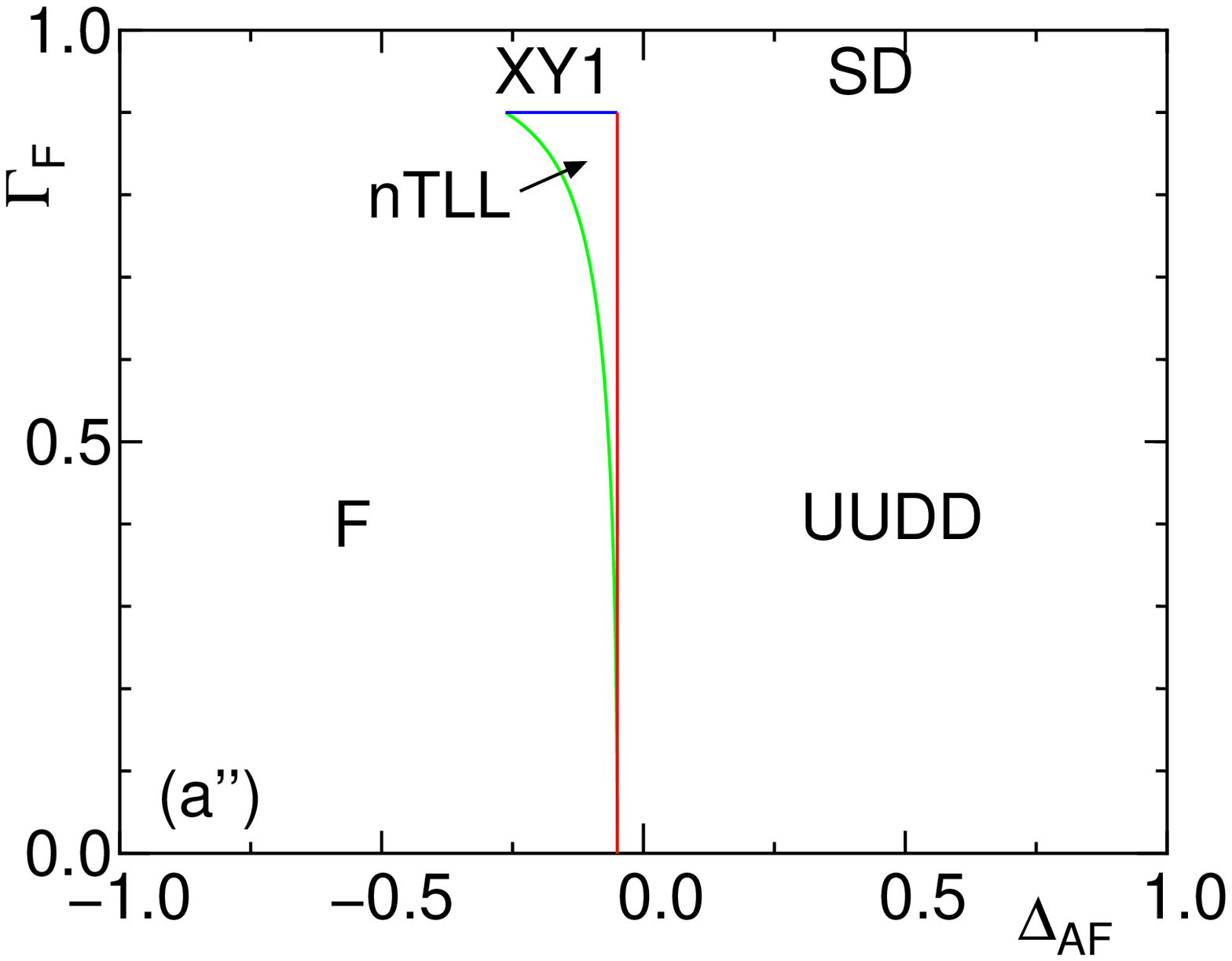}
   \end{center}
   \begin{center}
      \includegraphics[scale=0.29]{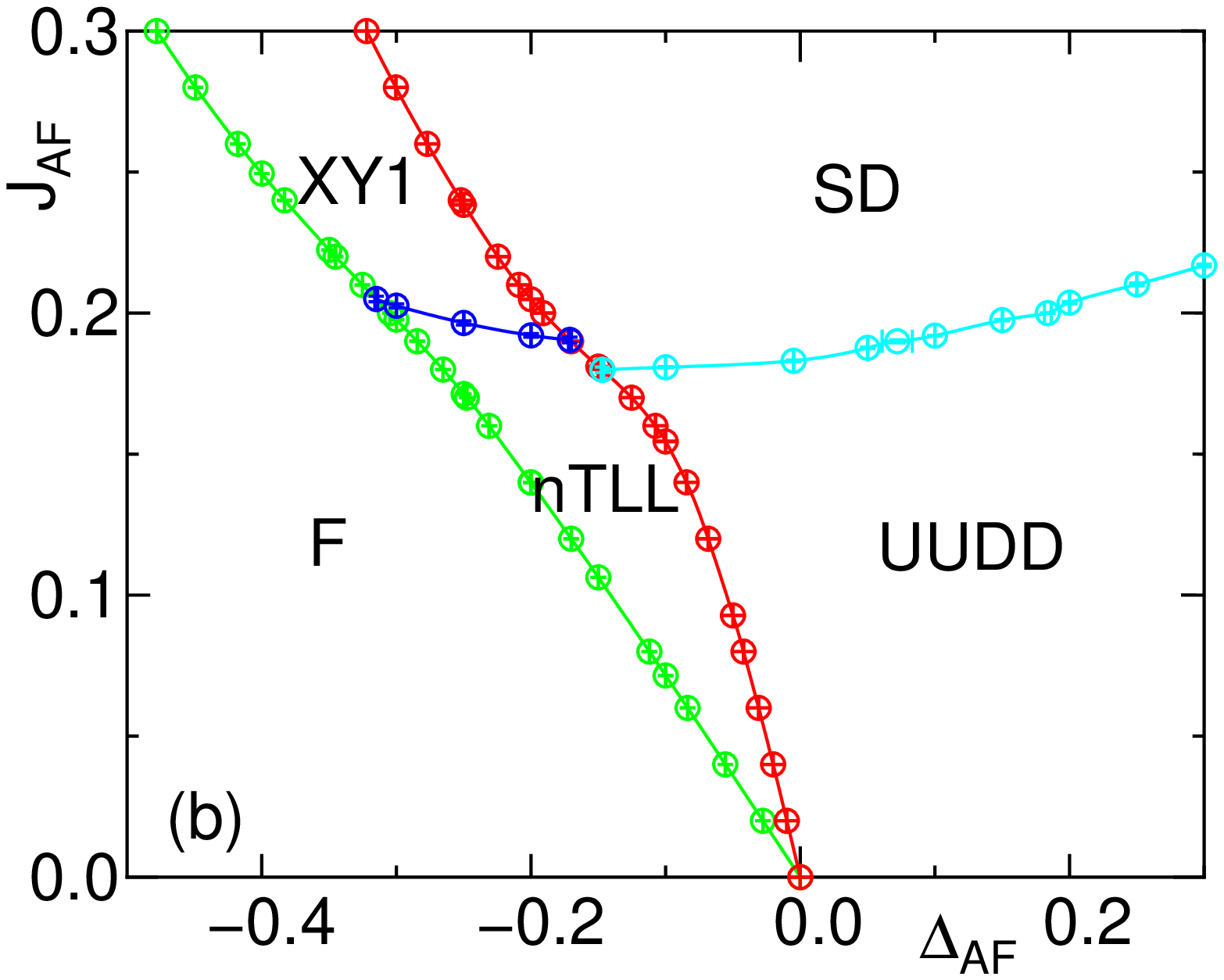}~~
      \includegraphics[scale=0.29]{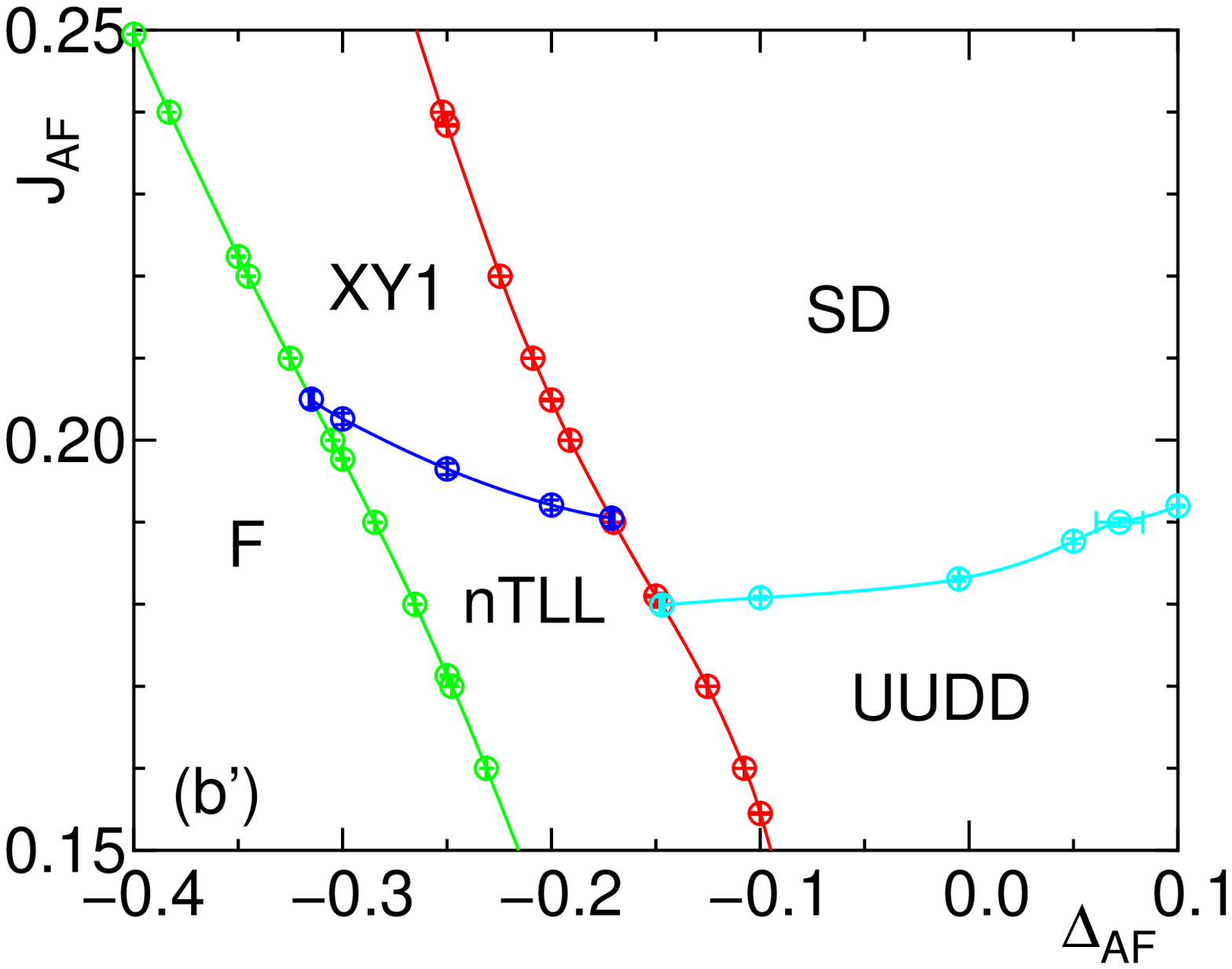}~~
      \includegraphics[scale=0.29]{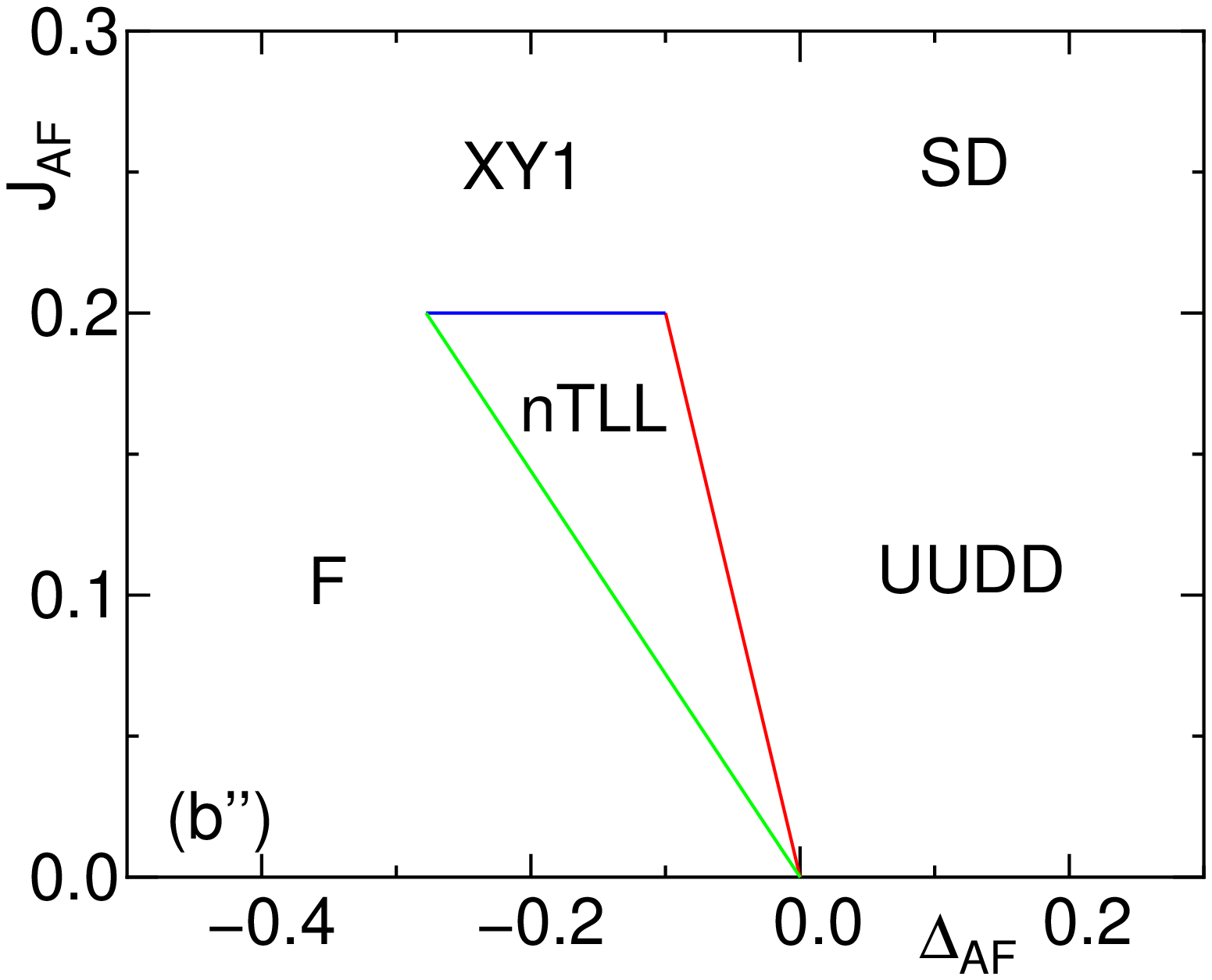}
   \end{center}
   \begin{center}
      \includegraphics[scale=0.29]{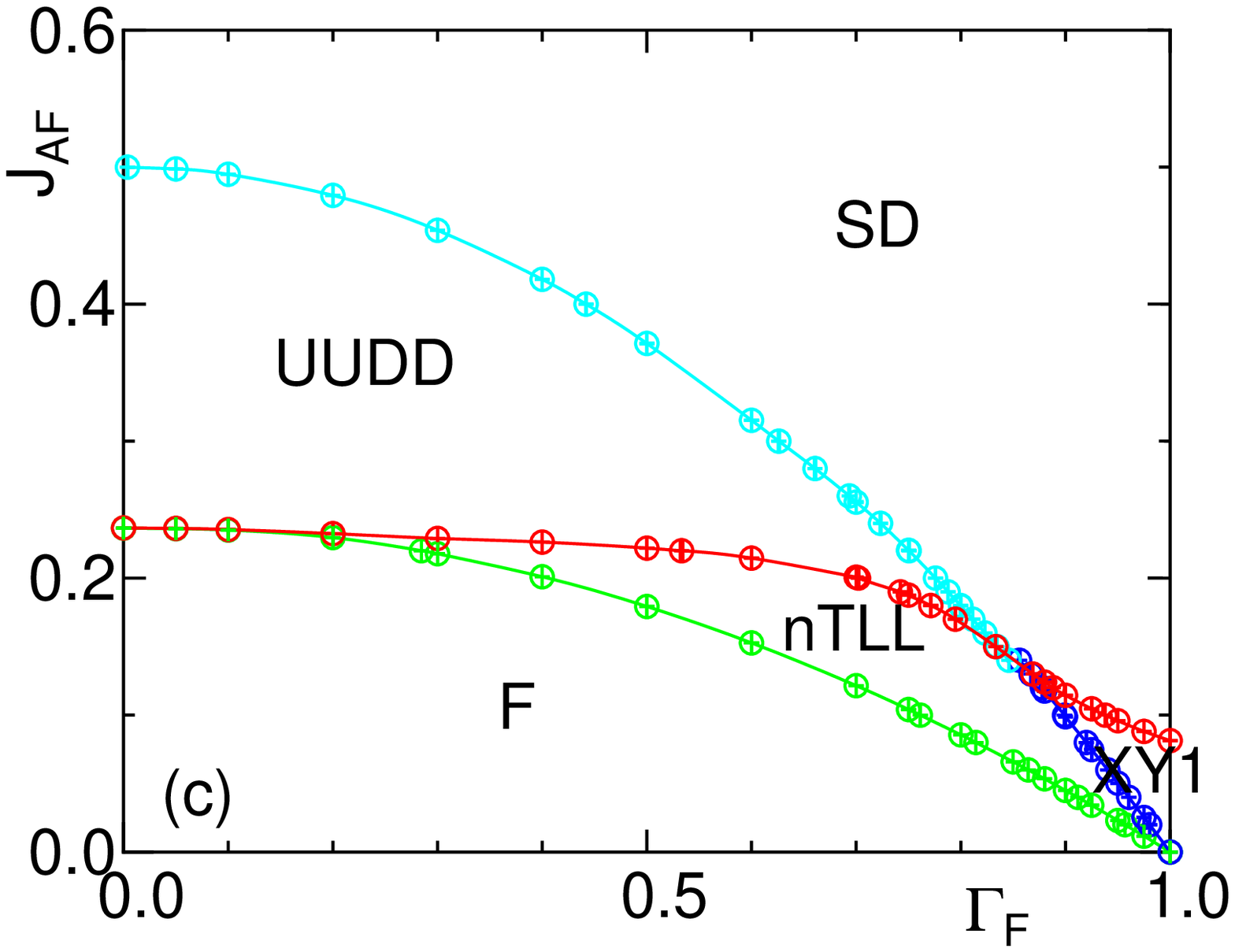}~~
      \includegraphics[scale=0.29]{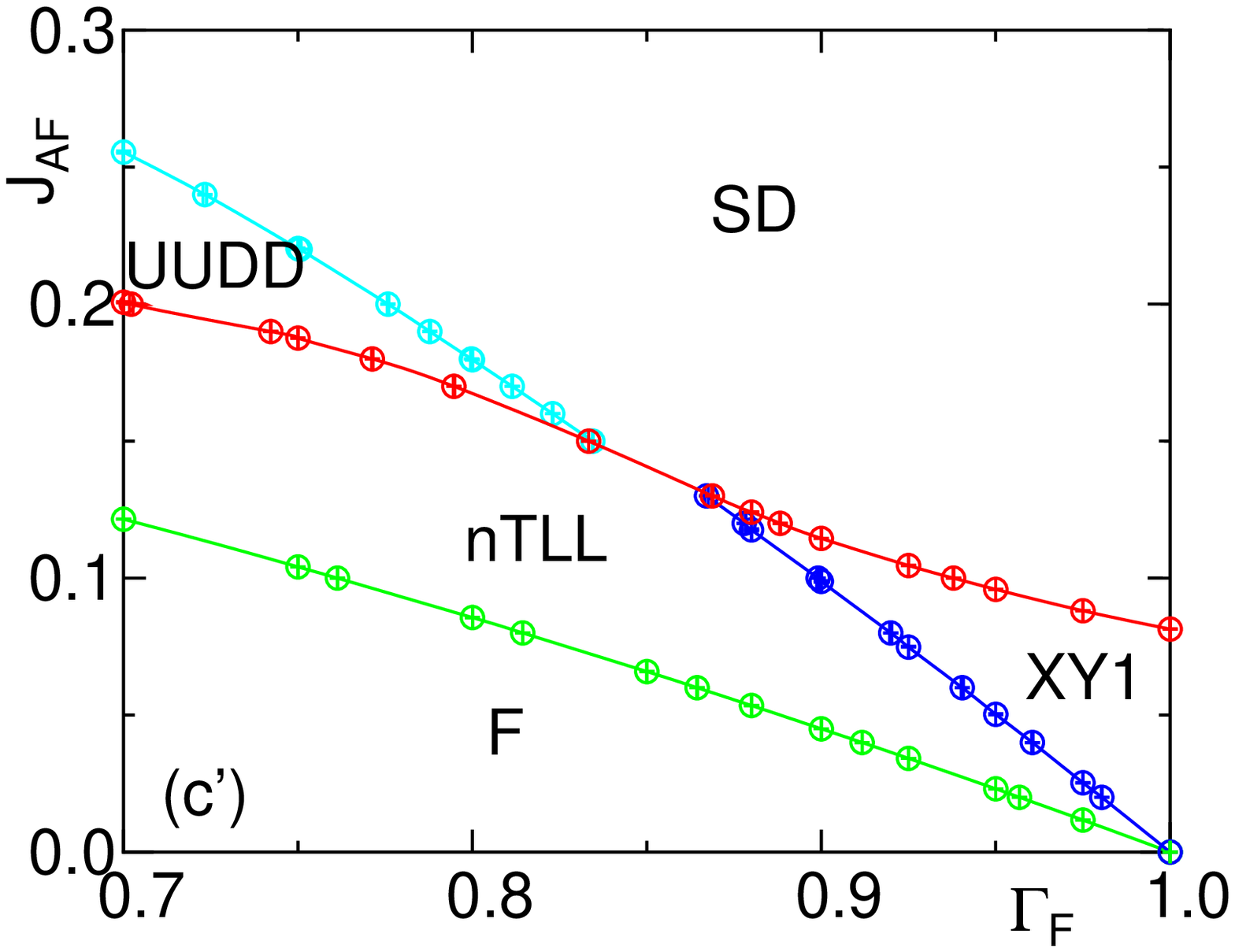}~~
      \includegraphics[scale=0.29]{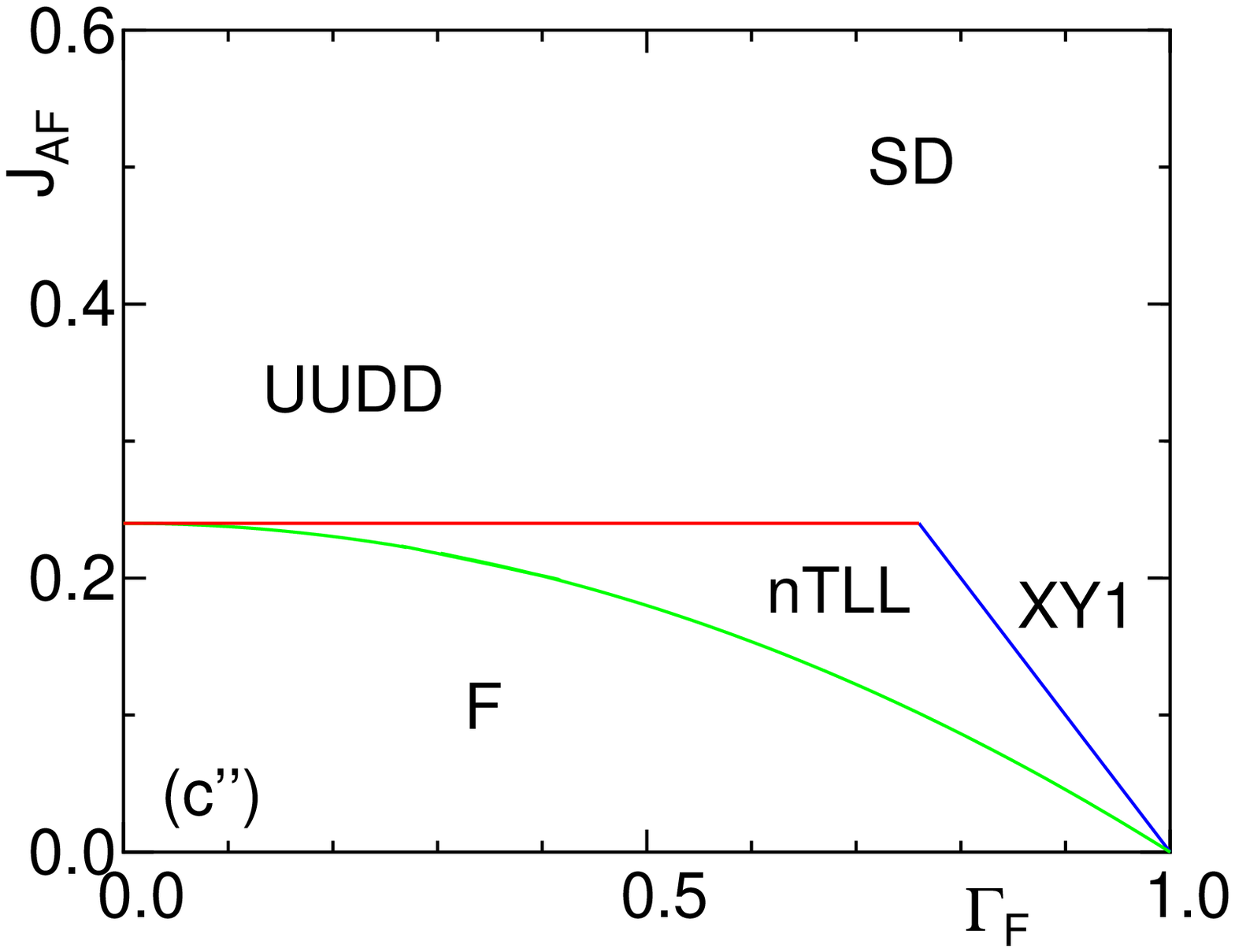}
   \end{center}
   \vskip-1cm
   \caption{Ground-state phase diagrams (a)~on the $\Delta_{\rm AF}$ versus
            $\Gamma_{\rm F}$ plane in the case of \hbox{$J_{{\rm F}}\!=\!1.0$}
            and \hbox{$J_{{\rm AF}}\!=\!0.1$}, (b)~on the $\Delta_{\rm AF}$
            versus $J_{\rm AF}$ plane in the case of
            \hbox{$J_{{\rm F}}\!=\!1.0$} and
            \hbox{$\Gamma_{{\rm F}}\!=\!0.8$}, and (c)~on the
            $\Gamma_{\rm F}$ versus $J_{\rm AF}$ plane in the case of
            \hbox{$J_{{\rm F}}\!=\!1.0$} and
            \hbox{$\Delta_{{\rm AF}}\!=\!-0.12$}, determined numerically in
            the present work.  In (a'), (b'), and (c'), parts of (a), (b), and
            (c) are enlarged, respectively.
            The results of the perturbation calculations are shown in (a''),
            (b''), and (c''); by these calculations the $XY$1-SD, SD-UUDD, and
            F-$XY1$ phase transition lines cannot be obtained.}
   \label{fig:fig1}
\end{figure}

\begin{figure}[ht]
   \vskip10cm
   \begin{center}
      \includegraphics[scale=0.45]{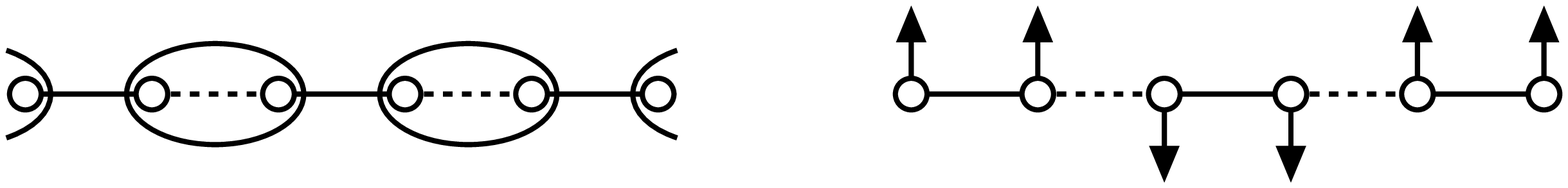}
   \end{center}
   \vskip-8cm
   \caption{Physical pictures of the SD (left) and UUDD (right) states.
            Open circles denote \hbox{$S\!=\!1/2$} spins, and solid and dotted
            lines denote ferromagnetic and antiferromagnetic bonds, 
            respectively.  Ellipses represent singlet pairs of two
            \hbox{$S\!=\!1/2$} spins, while arrows denote fixed projection
            values of the \hbox{$S\!=\!1/2$} spins.
            As can be seen from these figures, the SD phase is a unique gapped phase, while the UUDD phase is
            a doubly degenerate gapped phase.}
   \label{fig:fig2}
\end{figure}

First of all, we show in Fig.~\ref{fig:fig1} our numerical results of the
ground-state phase diagrams obtained in the following cases: (a)~the phase
diagram on the $\Delta_{\rm AF}$ versus $\Gamma_{\rm F}$ plane in the case of
\hbox{$J_{{\rm F}}\!=\!1.0$} and \hbox{$J_{{\rm AF}}\!=\!0.1$}, (b)~that
on the $\Delta_{\rm AF}$ versus $J_{\rm AF}$ plane in the case of
\hbox{$J_{{\rm F}}\!=\!1.0$} and \hbox{$\Gamma_{{\rm F}}\!=\!0.8$}, and
(c)~that on the $\Gamma_{\rm F}$ versus $J_{\rm AF}$ plane in the case of
\hbox{$J_{{\rm F}}\!=\!1.0$} and \hbox{$\Delta_{{\rm AF}}\!=\!-0.12$}.
These diagrams contain the ferromagnetic (F), $XY$1, singlet-dimer (SD),
and up-up-down-down (UUDD) phases in addition to the nTLL phase which appears
in wide regions.  The physical pictures of the SD and UUDD states are sketched
in Fig.~\ref{fig:fig2}.
The SD state is a unique and symmetry protected topological gapped state
and the UUDD state is a doubly degenerate gapped state.

Let us now discuss how to determine numerically the phase boundary lines in
the phase diagram shown in Fig.~\ref{fig:fig1}.  We denote, respectively, by
$E_0^{\rm P}(N,M)$ and $E_1^{\rm P}(N,M)$, the lowest and second-lowest energy
eigenvalues of the Hamiltonian~${\cal H}$ under periodic boundary conditions
within the subspace of fixed $N$ and $M$, where \hbox{$M(=\!0, \pm 1, \cdots,
\pm N/2)$} is the total magnetization.   Furthermore, we denote by
$E_0^{\rm T}(N,M,P)$ the lowest eigenvalue of ${\cal H}$ under twisted
boundary conditions within the subspace of fixed $N$, $M$, and $P$, where
\hbox{$P(=\!\pm 1)$} is the eigenvalue of the space inversion operator with
respect to the twisted bond.  We have numerically calculated these energies
for finite-size systems with up to \hbox{$N\!=\!28$} spins by means of the
exact-diagonalization method.  The ground-state energy of the finite-$N$ 
system is given by $E_0^{\rm P}(N,N/2)$ in the F phase and by
$E_0^{\rm P}(N,0)$ in the other phases.
In the following way, we have
estimated the finite-size critical values of the interaction parameters  for
each phase transition.   Then, the phase boundary line for the transition has
been obtained by connecting the results for the \hbox{$N\!\to\!\infty$}
extrapolation of the finite-size critical values.

First, the phase transition between the $XY$1 and SD phases is the
Berezinskii-Kosterlitz-Thouless (BKT) transition\cite{BKT,giamarchi}.  It is
known that for this transition, the level spectroscopy method developed by
Nomura and Kitazawa\cite{LSNK} is very powerful for calculating the phase
transition  line.  This method implies that the finite-size critical values
are estimated from
\begin{equation}
    \Delta E_0^{\rm P}(N,2) = \Delta E_0^{\rm T}(N,0,-1),
    \label{eq:ls}
\end{equation}
under the condition that $\Delta E_0^{\rm T}(N,0,+1)$ is larger than these
excitations,
where $\Delta E_0^{\rm P}(N,M) = E_0^{\rm P}(N,M) - E_0^{\rm P}(N,0)$ and
$\Delta E_0^{\rm T}(N,M,P) = E_0^{\rm T}(N,M,P) - E_0^{\rm P}(N,0)$.
This equation is also applicable to estimate the nTLL-UUDD phase transition
line (see Appendix).
Secondly, the phase transition between the SD and UUDD phases is the 2D
Ising-type transition.  Therefore, as is well known, the phase transition line
is determined by the phenomenological renormalization-group (PRG)
method\cite{PRG}.  Then, to estimate the finite-size critical values, we solve
the PRG equation,
\begin{equation}
   N\,\Delta E_1^\P(N,0) = (N+4)\,\Delta E_1^\P(N+4,0)\,,
\end{equation}
where $\Delta E_1^\P(N,M) \equiv E_1^\P(N,M) - E_0^\P(N,M)$.
Thirdly, as for the phase transition between the nTLL and $XY$1 phases,
the nTLL state accompanies two-magnon bound-states, while the $XY$1
state does not.  Therefore, in the ground-state magnetization curve for the
finite-size system, the magnetization increases from \hbox{$M\!=\!0$} to
\hbox{$M\!=\!2$} in the former state and from \hbox{$M\!=\!0$} to
\hbox{$M\!=\!1$} in the latter state.  Thus, the finite-size critical values
are estimated from
\begin{equation}
   \Delta E_0^\P(N,2) = 2\,\Delta E_0^\P(N,1)\,.
   \label{eq:xy1-ntll}
\end{equation}
We note that the binding energy of two magnons\cite{tonegawa-1} is defined by
\hbox{$E_{\rm bind}(N)\!\equiv\!
\Delta E_0^\P(N,2)\!-\! 2\,\Delta E_0^\P(N,1)$}.
Accordingly, Eq.(\ref{eq:xy1-ntll}) is also the condition of
\hbox{$E_{\rm bind}(N)\!=\!0$}.
Lastly, it is apparent that the finite-size critical values for the
phase transitions between the F phase and one of the $XY$1 and nTLL phases
are estimated from 
\begin{equation}
   E_0^{\rm P}(N,N/2)\!=\!E_0^{\rm P}(N,0)\,.
\end{equation}

\section{Perturbation Theory}

In the nTLL phase, the important states of two \hbox{$S\!=\!1/2$} spins
connected by the Ising-like ferromagnetic coupling $J_{\rm F}$
are $| \uparrow \uparrow \rangle$ and $| \downarrow \downarrow \rangle$,
and the remaining two states 
$|(1/\sqrt{2}) (\uparrow \downarrow \pm \downarrow \uparrow) \rangle$ have
higher energies.  To describe the nTLL state, we introduce a pseudo-spin
operator \mbox{\boldmath $T$} with \hbox{$T\!=\!1/2$},
where $| T^z =1/2 \rangle = | \uparrow \uparrow \rangle$
and $| T^z =-1/2 \rangle = | \downarrow \downarrow \rangle$.
We perturbationally derive the effective Hamiltonian $\cal H_\eff$
described by \mbox{\boldmath $T$}.
We take the first term of the right-hand side of Eq.(\ref{eq:hamiltonian}) as
the unperturbed Hamiltonian, and the second term as the perturbation.
Up to the second order perturbation calculation,
we obtain the following $XXZ$ model, apart from the energy
shift,
\begin{equation}
    {\cal H}_\eff
    = \sum_{j=1}^{N/2} 
      \left\{ J_\eff^\perp (T_j^x T_{j+1}^x + T_j^y T_{j+1}^y) 
              + J_\eff^z T_j^z T_{j+1}^z \right\}
    \label{eq:heff}
\end{equation}
with
\begin{eqnarray}
    &&J_\eff^\perp = c_1 - 2c_2 + c_3,~~~~~J_\eff^z = b+ c_1 + 2c_2 + c_3,  \\
    &&b = J_\AF \Delta _\AF,~~~~~
      c_1 = {J_\AF^2 \over 8J_\F(1-\Gamma_\F)},~~~~~
      c_2 = {J_\AF^2 \over 8J_\F},~~~~~
      c_3 = {J_\AF^2 \over 8J_\F(1+\Gamma_\F)},
\end{eqnarray}
where $J_\eff^\perp \ge 0$.
The very well known exact solution\cite{giamarchi} for Eq.~(6) shows that
the ground state phase diagram of ${\cal H}_\eff$
consists of the F phase, the TLL phase, and the N\'eel phase.
These phases correspond, respectively, to the F phase, the nTLL
phase and the UUDD phase of the original model~(\ref{eq:hamiltonian}).
Thus, the F-nTLL boundary line of the original model~(\ref{eq:hamiltonian}) is
determined by $J_\eff^z = -J_\eff^\perp$,
while the nTLL-UUDD boundary line by $J_\eff^z = J_\eff^\perp$.
We note that the same effective Hamiltonian with the periodic boundary
condition is obtained irrespective of the
periodic boundary condition or the twisted boundary condition of the original
Hamiltonian~(\ref{eq:hamiltonian}).

The nTLL-$XY1$ boundary can be estimated by considering the energy cost of
replacing a $|T^z = \pm1/2 \rangle$ pseudo-spin in the ${\cal H}_\eff$ picture
by $|(1/\sqrt{2}) (\uparrow \downarrow + \downarrow \uparrow) \rangle$.
If this energy cost is negative,
a macroscopic number of spin pair states with 
$|(1/\sqrt{2}) (\uparrow \downarrow + \downarrow \uparrow) \rangle$
are generated, which brings about the breakdown of the ${\cal H}_\eff$ picture.
On the other hand, if this energy cost is positive,
the $|(1/\sqrt{2}) (\uparrow \downarrow + \downarrow \uparrow) \rangle$ spin
pair state are scarcely yielded,
resulting in the stability of the ${\cal H}_\eff$ picture.
The replacement of a pseudo-spin
$|T^z = -1/2\rangle \Rightarrow |(1/\sqrt{2}) (\uparrow \downarrow
+ \downarrow \uparrow) \rangle$
in the $M=0$ ground state leads to the $M=1$ state.
Thus, if this cost is positive, the $M=1$ excitation is gapped,
which is consistent with the picture of the nTLL state.
We note that the replacement of a $|T^z = -1/2 \rangle$ spin state by a
$|T^z = +1/2 \rangle$ one brings about the $M=2$ state.
After some calculations, we obtain this energy cost as 
$(1/2)\{ J_\F(1-\Gamma_\F) - J_\AF\}$.
Thus, the nTLL-$XY1$ boundary line is estimated by $J_\AF = J_\F(1-\Gamma_\F)$.

The above three boundary lines are depicted in Figs.~\ref{fig:fig1}(a''),
(b''), and (c'').

\section{Concluding Remarks}

We have determined the ground-state phase diagram of the \hbox{$S\!=\!1/2$}
ferromagnetic-anti\-ferro\-magnetic bond-alternating chain described by the
Hamiltonian of Eq.~(\ref{eq:hamiltonian}) by using mainly the numerical
methods.  Discussing the case where the ferromagnetic interactions with 
the Ising-type anisotropy are much stronger than the antiferromagnetic ones
with the $XY$-type anisotropy, we have shown that in the phase diagrams
(see Fig.~\ref{fig:fig1}), there appear the F, $XY$1, SD, and UUDD phases as
well as the nTLL phase.  We have also developed the perturbation theory, which
qualitatively explains the numerically obtained phase
boundary lines associated with the nTLL phase.
This paper gives, in a succession of our previous
work\cite{tonegawa-1}, a report of the appearance of the nTLL phase
in an unfrustrated \hbox{$S\!=\!1/2$} chain under no external magnetic field.

It should be noted that the SDW$_2$\cite{hikihara-1} ('SDW' is an abbreviation
for 'spin-density-wave') does not appear in the ground-state phase diagram of
the present model.  The reason for this is as follows.
In both of the nTLL state and the SDW$_2$ state,
both of the nematic four-spin correlation 
$C_2(j) \equiv \langle S_1^+ S_2^+ S_{1+j}^- S_{2+j}^- \rangle$ and
the longitudinal two-spin correlation 
$C_z(j) \equiv \langle S_1^z S_{1+j}^z \rangle$, where $j$ is even, decay in
power laws.  The former correlation is dominant in the nTLL state,
while the latter is dominant in the SDW$_2$ state.
Namely, $\eta_2 < \eta_z$ in the nTLL state, whereas $\eta_2 > \eta_z$ in the
SDW$_2$ state, where $\eta_2$ and $\eta_z$ are the power-decay exponents of
$C_2(j)$ and $C_z(j)$, respectively.
Since there is the TLL relation $\eta_2 \eta_z = 1$,
it holds $\eta_2 > 1$ in the SDW$_2$ state.
However, this condition $\eta_2 > 1$ is nothing but the necessary condition
for the UUDD state (see Appendix).
In the zero magnetization case,
the operator coming from the Umklapp process becomes relevant due to
$\eta_2 > 1$, 
which leads to the UUDD state.
Thus, the SDW$_2$ region does not exist in the ground state phase diagram.
On the other hand, in the finite magnetization case,
since the operator induced by the Umklapp process does not exist,
the SDW$_2$ region appears in the phase diagram.

One can see several tricritical points in the phase diagrams given in
Fig.~\ref{fig:fig1}; these are the F-$XY$1-nTLL, $XY$1-SD-UUDD, and
$XY$1-UUDD-nTLL tricritical points in the case~(a), the F-$XY$1-nTLL,
$XY$1-SD-nTLL, and SD-UUDD-nTLL tricritical points in the case~(b),
and the SD-UUDD-nTLL and $XY$1-SD-nTLL tricritical points in the case~(c).
According to the discussion on the university class (see Appendix), on the
other hand, two tricritical points associated with the $XY$1, SD, UUDD,
and nTLL phases should merge into one $XY$1-SD-UUDD-nTLL tetracritical
point. In order to obtain numerically this tetracritical point, it is
indispensable to develop a new numerical method which is applicable commonly
to the phase transition between the $XY1$ and nTLL phases and to that between
the SD and UUDD phase.  Since both of these transitions are of the 2D
Ising-type, it seems not to be impossible to find this method.  However,
it is has not yet been succeeded at present, and thus this problem is beyond
the scope of the present study.  Furthermore, it may be expected that the nTLL
phase appear even in the case where the antiferromagnetic interactions are
stronger than the ferromagnetic interactions in the ground-state phase diagram
of the present system, if
\hbox{$1.0\gg\Gamma_{{\rm F}}\!\geq\!0.0.$} and
\hbox{$1.0\gg|\Delta_{{\rm AF}}|$}.
We are now planning to perform this calculation, and the results will be
reported in the near future.

\section*{Acknowledgments}

This work has been partly supported by JSPS KAKENHI, Grant Numbers 16K05419,
16H01080 (J-Physics), 18H04330 (J-Physics), JP20K03866, JP20H05274, and 21H05021.
We also thank the Supercomputer Center, Institute for Solid State Physics,
University of Tokyo and the Computer Room, Yukawa Institute for Theoretical
Physics, Kyoto University for computational facilities.

\appendix

\section{The Level Spectroscopy Method for the nTLL-UUDD Transition}

Let us consider the transverse two-spin correlation
$C_1(j) \equiv\langle S_1^+ S_{1+j}^- \rangle$ and the nematic four-spin
correlation
$C_2(j) \equiv \langle S_1^+ S_2^+ S_{1+j}^- S_{2+j}^- \rangle$, where $j$ is
even.  Both of them decay in power laws in the $XY1$ phase with the relation
$\eta_2 = 4\eta_1$ \cite{tonegawa-1}, where $\eta_1$ and $\eta_2$ are the
power-decay exponents of $C_1(j)$ and $C_2(j)$, respectively.
As is well known, the BKT transition from the $XY1$ phase to the unique gapped
phase (the SD phase in the present case)
occurs at $\eta_1=1/4$ (hence $\eta_2=1$),
whereas to the doubly degenerate gapped phase (the UUDD phase in the present
case) at $\eta_1=1$ (accordingly $\eta_2=4$) \cite{KNO,LSNK}.
In the level spectroscopy method by Nomura and Kitazawa\cite{LSNK} for the BKT
transition between the $XY1$ phase and the SD phase,
we search the $\eta_2=1$ line.
This is clear from the fact that the excitation $\Delta E_0^{\rm P}(N,2)$ is
used in Eq.(\ref{eq:ls}),
since $\Delta E_0^{\rm P}(N,2)$ is closely related to $\eta_2$ as
$\eta_2 = \lim_{N \to \infty} [N\Delta E_0^{\rm P}(N,2) / \pi v_{\rm s}]$,
where $v_{\rm s}$ is the spin wave velocity \cite{KNO,LSNK,cardy,sakai1998}.

On the other hand, in the nTLL phase,
$C_1(j)$ exhibits the exponential decay, while $C_2(j)$ the power decay.
Thus, the role of $\eta_1$ in the BKT transitions from the $XY1$ phase is
superseded by $\eta_2$ in those from the nTLL phase.
Namely, the BKT transition from the nTLL phase to the unique gapped phase
occurs at $\eta_2 =1/4$,
while that to the doubly degenerate gapped state at $\eta_2=1$.
Hence, the BKT transition from the nTLL phase to the doubly degenerate gapped
phase (the UUDD phase in the present case) can also be determined by
Eq.(\ref{eq:ls}) in which the line $\eta_2=1$ is searched.

Neither the BKT transition from the $XY1$ phase to the UUDD phase
nor that from the nTLL phase to the SD phase
occurs on the $\eta_2=1$ line.
This is because the former transition should occur at $\eta_1=1$ ($\eta_2=4$) even if it occurs.
Similarly the latter should occur at $\eta_2 = 1/4$ even if it occurs.
Accordingly there should exist the $XY$1-SD-UUDD-nTLL tetracritical point on
the $\eta_2=1$ line which is determined by Eq.(\ref{eq:ls}).
A similar discussion can also be applied to the \hbox{$S\!=\!1$} chain with the
XXZ and on-site anisotropies which has been discussed, for example, by Chen et
al. \cite{chen}.  Namely, the $XY$1-Haldane-N{\'e}el-nTLL tetracritical point should
exist in this \hbox{$S\!=\!1$} chain.


\end{document}